\documentclass[twocolumn]{IEEEtran}

\usepackage{graphicx}

\input{tcilatex}

\begin{document}

\title{Electron g-factor Engineering in III-V Semiconductors for Quantum
Communications}
\author{Hideo Kosaka*$^{1}$, Andrey A. Kiselev$^{2}$, Filipp A. Baron$^{1}$, Ki Wook
Kim$^{2}$ and Eli Yablonovitch$^{1}$ \\
$^{1}$University of California, Los Angeles, Electrical Engineering Dept.,\\
Los Angeles, California 90024, USA\\
$^{2}$Department of Electrical and Computer Engineering, North Carolina\\
State University, Raleigh, North Carolina 27695, USA\thanks{$^{\ast }$ Hideo
Kosaka is on leave from NEC, Fundamental Research Laboratories, 34,
Miyukigaoka, Tsukuba 305-8501, Japan.}}
\maketitle

\begin{abstract}
An entanglement-preserving photo-detector converts photon polarization to
electron spin. Up and down spin must respond equally to oppositely polarized
photons, creating a requirement for degenerate spin energies, g$_{e}\approx $%
0 for electrons. We present a plot of g$_{e}$-factor versus lattice
constant, analogous to bandgap versus lattice constant, that can be used for
g-factor engineering of III-V alloys and quantum wells
\end{abstract}

The major motive for the research on III-V semiconductors has been
the development of opto-electronic devices for optical
communications. One of the key inventions for improved
semiconductor lasers was electronic band structure engineering
based on strained heterostructures \label{Yab88}\cite {Yab88}.
Practical realization of quantum communications \cite
{Bennett89,Bennett92} is expected to require
entanglement-preserving photo-detectors in which quantum
information is transmitted by photon polarization through an
optical fiber, and then transferred to electron spin in a
semiconductor \cite{Vrijen00,Vrijen??}.
\begin{figure}
\includegraphics[width=9.2cm]{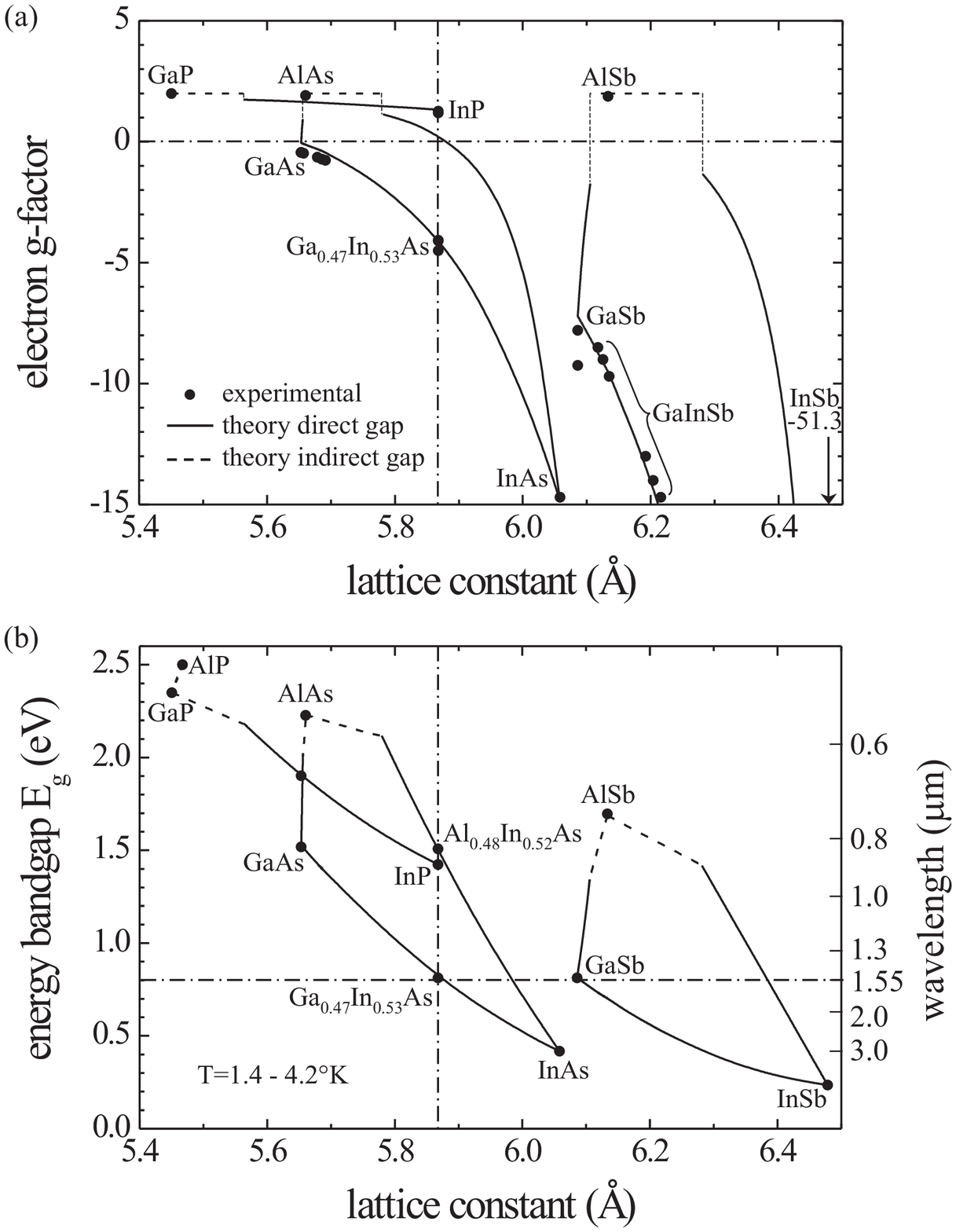}
\caption{(a) A graph of g$_{e}$-factors for conduction electrons
in III-V semiconductors as a function of lattice constant. Dots show experimental g$%
_{e}$-factors, solid curves show direct bandgap materials, and
dashed curves show indirect bandgap materials. The vertical
dash-dotted line indicates the lattice constant of bulk InP, that
is normally used for optical communication devices. Bulk
g$_{e}$-factors are plotted for direct bandgap materials (InP
[7,8], GaAs [9], GaInAs [10-12], InAs [13], GaSb [14,15], GaInSb
[16] and InSb [17]), and defect or impurity related g-factors are
plotted for indirect bandgap materials (GaP [18], AlAs [19], and
AlSb [20]). All data were taken at low temperature 1.4 -
4.2$^\circ$K, but Ref. 13 is taken at 30$^\circ$ K.   (b) Energy
bandgaps for the materials shown in Fig. 1(a) at temperatures
between 1.4 and 4.2$^\circ$ K. }
\end{figure}
To maintain the entanglement, the photo-detector should absorb
equally into up and
down electron spin states, and thus the electron g-factor should be engineered for g$%
_{e}\approx $0.Fortunately, the familiar band structure engineering of
effective mass can equally well control g-factor as well. There are
additional requirements in Ref. 5 for moderately long spin coherence times,
and for a hole g-factor $\left| g_{h}\right| $\TEXTsymbol{>}\TEXTsymbol{>}0,
large enough to lift the Kramers' degeneracy of the valence band. Various
III-V alloys and quantum wells can be engineered to produce the right
g-factor combinations.

To commence the task of g-factor engineering, we have graphed the
experimental electron g$_{e}$-factors in III-V semiconductors as a
function of the lattice constant in Fig. 1(a). This is analogous
to the famous graph of bandgaps versus lattice constant plotted in
Fig. 1(b). In these figures, we have inscribed vertical \&
horizontal dash-dotted lines to illustrate design preferences. The
horizontal axis in Fig. 1(a) shows a requirement for zero electron
g$_{e}$-factor, g$_{e}$ $\approx $0. The horizontal
dash-dotted line in Fig. 1(b), is a preference for a bandgap of 0.8 eV, or $%
\lambda $=1.55 $\mu $m, corresponding to the optimum wavelength
for fiber-optic communications, but shorter wavelengths $\lambda
$=1.3 $\mu $m are also acceptable for quantum communication. To
lattice match the entanglement-preserving photo-detector to a
conventional InP substrate, a vertical dash-dotted line is shown
at lattice constant a = 5.86 \AA .

To fill in the gaps between the experimental g$_{e}$-factor points
in Fig. 1(a), some theoretical curves were included in Fig. 1(a).
The following simple formula derived from the k$\cdot $p
perturbation theory \cite{Roth59} was used for the theoretical
curves:

$\bigskip $%
\begin{equation}
g_{e}=2-\frac{2}{3}\frac{E_{p}\Delta }{E_{g}(E_{g}+\Delta )},  \label{eq1}
\end{equation}

\bigskip

where $E_{g}$ is the energy bandgap, $\Delta $ is the spin-orbit splitting
energy, and $E_{p}$ is the energy equivalent of the principal interband
momentum matrix element. We used experimentally determined $E_{g}$, and
linearly interpolated $\Delta $ and $E_{p}$ for alloys, utilizing known
values for pure compounds. No fitting parameters were used. Although the
formula includes only the lowest conduction band, the highest valence band,
and a spin split-off valence band, it agrees reasonably well with the
experimental data. Dotted lines in the figure show the range of indirect
material, where Eq. (1) is not applicable, and g$_{e,indirect}\approx $2.

The trend in both Fig. 1(a) \& 1(b) is for g$_{e}$-factor be more negative
as the bandgap drops. The main exception is InP. InP is the only III-V to
show positive g$_{e}$-factor at moderate bandgaps. This is because of the
remarkably small spin-orbit splitting $\Delta $ = 0.108 eV in InP.
Considering only bulk III-V semiconductors, one prefers to reduce $E_{g}$
down to 0.8 eV (1.55 $\mu $m), while keeping $\Delta $ no larger than 0.11
eV, in order to achieve near-zero g$_{e}$-factor ($E_{p}$ is always around
22 eV). No bulk material is readily available with these characteristics.
Much more design freedom can be achieved in multilayer heterostructures. For
a reliable quantitative analysis to establish a proper design, it will be
necessary to take into account the effects of quantum confinement as well as
the strain-induced valence band splitting to fulfill the electron \& hole
g-factor requirements of an entanglement-preserving photo-detector.

Acknowledgement: The project or effort depicted is sponsored by the Defense
Advanced Research Projects Agency \& Army Research Office projects
MDA972-99-1-0017, DAAD19-00-1-0172, and MDA972-00-1-0035. The content of the
information does not necessarily reflect the position or the policy of the
government, and no official endorsement should be inferred.

\bigskip

\bibliographystyle{IEEEtr}

\begin{thebibliography}{99}
\bibitem{Yab88}  Yablonovitch, E. and Kane, E.O.: 'Band structure
engineering of semiconductor lasers for optical communications', J.
Lightwave Technol., 1988, 6, (8), pp. 1292-1299

\bibitem{Bennett89}  Bennett, C.H. and Brassard, G.: 'The dawn of a new era
for quantum cryptography; The experimental prototype is working!', SIGACT
News, 1989, 20, pp. 78-82

\bibitem{Bennett92}  Bennett, C.H. and Wiesner, S.J.: 'Communication via
one- and two-particle operators on Einstein-Podolsky-Rosen states', Phys.
Rev. Lett., 1992, 69, pp. 2881-2884

\bibitem{Vrijen00}  Vrijen, R. et al.: 'Electron spin resonance transistors
for quantum computing in Silicon-Germanium hetero-structures', Phys. Rev. A,
2000, 62, (1), pp. 012306-1-10 \ (http://xxx.lanl.gov/abs/quant-ph/9905096)

\bibitem{Vrijen??}  Vrijen, R. and Yablonovitch, E.: 'A solid-state spin
coherent photo-detector for quantum communication', Physica E, to be
published\ \ (http://arxiv.org/abs/quant-ph/0004078)

\bibitem{Roth59}  Roth, L.M., Lax, B., and Zwerdling, S.: 'Theory of optical
magneto-absorption effects in semiconductors', Phys. Rev., 1959, 114, (1),
pp. 90-100

\bibitem{Weisbuch75}  Weisbuch, C., Herrmann, C.: 'Optical detection of
conduction electron spin resonance in InP', Solid State Commun., 1975, 16,
(5), pp. 659-661

\bibitem{Oestreich96}  Oestreich, M., Hallstein, S., Heberle, A.P., Eberl,
L., Bauser, E., and Ruhle, W.W.; 'Temperature and density dependence of the
electron Lande g factor in semiconductors', Phys. Rev. B, 1996, 53, (12),
pp. 7911-7916

\bibitem{White72}  White, A.A., Hinchlifee, J., Dean, P.J.: 'Zeeman spectra
of the principal bound exciton in Sn-doped gallium arsenide', Solid State
Commun., 1972, 10, (6), p. 497-500

\bibitem{Weisbuch77}  Weisbuch, C., Herrmann, C.: 'Optical detection of
conduction-electron spin resonance in GaAs, Ga1-xInxAs, and Ga1-xAlxAs',
Phys. Rev. B, 1977, 15, (2), p. 816-822

\bibitem{Kowalski96}  Kowalski, B., Omling, P., Meyer, B.K., Hofmann, D.M.,
H\"{a}rle, V., Scholz, F., and Sobkowicz, P.,: 'Optically detected spin
resonance of conduction band electrons in InGaAs/InP quantum wells',
Semicond. Sci. Technol., 1996, 11, pp. 1416-1423

\bibitem{Dobers89}  Dobers, M., Vieren, J.P., and Guldner, Y.:
'Electron-spin resonance of the two-dimensional electron gas in
Ga0.47In0.53As-InP heterostructures', Phys. Rev. B, 1989, 40, (11), pp.
8075-8078

\bibitem{Konopka67}  Konopka, J.: Phys. Lett., 1967, 26A, p. 21

\bibitem{Herrmann77}  Herrmann, C., Weisbuch, C.: 'k$\cdot$p perturbation
theory in III-V compounds and alloys: a reexamination ', Phys. Rev. B, 1977,
15, (2), p. 823-833

\bibitem{Reine72}  Reine, M., Aggarwal, R.L., Lax, B.: 'Stress-modulated
magnetoreflectivity of gallium antimonide and gallium arsenide', Phys. Rev.
B, 1972, 5, (8), p. 3033-3049

\bibitem{Roth78}  Roth, A.P., Fortin, E.: 'Interband magneto-optical study
of the In1-xGaxSb alloy system', Can. J. Phys., 1978, 56, (11), p. 1468-1475

\bibitem{Isaacson68}  Isaacson, R.A.: Phys. Rev., 1968, 169, p. 312

\bibitem{Kaufmann76}  Kaufmann, U., Schneider, J., Rauber, A.: 'ESR
detection of antisite lattice defects in GaP, CdSiP2 and ZnGeP2', Appl.
Phys. Lett., 1976, 29, (5), p. 312-313

\bibitem{Glaser91}  Glaser, E.R., Kennedy, T.A., Molnar, B., and Sillmon,
R.S.: 'Optically detected magnetic resonance of group-IV and group-VI
impurities in AlAs and AlxGa1-xAs with x\TEXTsymbol{>}0.35', Phys. Rev. B,
1991, 43, (18), pp. 14540-14556

\bibitem{Glaser99}  Glaser, E.R., Kennedy, T.A., Bennett, B.R., and
Shanabrook, B.V.: 'Strong emission from As monolayers in AlSb', Phys. Rev.
B, 1999-I, 59, (3), pp. 2240-2244

\bigskip \newpage
\end{thebibliography}

\end{document}